\documentclass[sigconf, 10pt]{acmart}
\usepackage{graphicx} 
\usepackage{xcolor}
\usepackage{mathtools}
\usepackage[skip=0pt]{caption}
\usepackage{subfigure}
\usepackage{url}
\usepackage{balance}
\usepackage{amsmath}
\usepackage{wrapfig}
\usepackage{natbib}

\AtBeginDocument{%
  \providecommand\BibTeX{{%
    \normalfont B\kern-0.5em{\scshape i\kern-0.25em b}\kern-0.8em\TeX}}}

\newcommand{\st}{\textsuperscript{\textparagraph}}
\newcommand{\tsec}{\textsuperscript{\textsection}}

\title{AraSync: Precision Time Synchronization\\in Rural Wireless Living Lab}

\acmYear{2024}\copyrightyear{2024}
\acmConference[ACM MobiCom '24]{The 30th Annual International Conference on Mobile Computing and Networking}{November 18--22, 2024}{Washington D.C., DC, USA}
\acmBooktitle{The 30th Annual International Conference on Mobile Computing and Networking (ACM MobiCom '24), November 18--22, 2024, Washington D.C., DC, USA}
\acmDOI{10.1145/3636534.3697318}
\acmISBN{979-8-4007-0489-5/24/11}

\begin{document}

\author{\large Md Nadim\tsec, Taimoor Ul Islam\tsec, Salil Reddy\st, Tianyi Zhang\tsec, Zhibo Meng\tsec, Reshal Afzal\tsec, Sarath Babu\tsec, \\\vspace{-1.5mm} Arsalan Ahmad\tsec, Daji Qiao\tsec, Anish Arora\st, and Hongwei Zhang\tsec}

\email{{nadim, tislam, tianyiz, zhibom, reshal, sarath4, aahmad, daji, hongwei}@iastate.edu, {reddy.441,arora.9}@osu.edu}

\renewcommand{\shortauthors}{Nadim et al.}
\renewcommand{\shorttitle}{AraSync: Precision Time Synchronization in Rural Wireless Living Lab}

\affiliation{%
  \institution{\tsec Iowa State University, \st Ohio State University}
  \country{}
}

\begin{abstract}

Time synchronization is a critical component in network operation and management, and it is also required by Ultra-Reliable, Low-Latency Communications~(URLLC) in next-generation wireless systems such as those of 5G, 6G, and Open RAN. In this context, we design and implement \textit{AraSync} as an end-to-end time synchronization system in the ARA wireless living lab to enable advanced wireless experiments and applications involving stringent time constraints. We make use of Precision Time Protocol~(PTP) at different levels to achieve synchronization accuracy in the order of nanoseconds. Along with fiber networks, \textit{AraSync} enables time synchronization across the AraHaul wireless x-haul network consisting of long-range, high-capacity mmWave and microwave links. In this paper, we present the detailed design and implementation of \textit{AraSync}, including its hardware and software components and the PTP network topology. Further, we experimentally characterize the performance of \textit{AraSync} from spatial and temporal dimensions. Our measurement and analysis of the clock offset and mean path delay show the impact of the wireless channel and weather conditions on the PTP synchronization accuracy. 

\end{abstract}

\ccsdesc[500]{Networks~Network measurement}
\ccsdesc[500]{Computer systems organization~Real-time system architecture}

\keywords{Precision Time Protocol (PTP), ARA Wireless Living Lab, AraSync, Wired-PTP, Wireless-PTP}

\maketitle

\section{Introduction}

Next-generation wireless networks are characterized by features to enable Ultra-Reliable, Low-Latency Communications~(URLLC). As per the requirements of 6G~\cite{9972784}, it is essential to ensure an end-to-end data-rate of hundreds of gigabits per second with guaranteed sub-millisecond latency and jitter. 
To meet such stringent low-latency requirements,
time synchronization between interacting network components is of utmost priority. 

Precision Time Protocol~(PTP)~\cite{1048550,4579760,9120376,9535437} and Network Time Protocol (NTP)~\cite{103043} are the most widely used time synchronization protocols in various networks. PTP is specifically designed for industrial environments that involve control systems and is ideal for distributed systems because of its minimal bandwidth requirements and low processing overhead. In addition to lower overhead, PTP offers precise clock synchronization compared to NTP~\cite{Arnold2023May_2}. Depending on the network topology, PTP can achieve synchronization accuracy in the order of sub-microsecond and down to sub-nanosecond range in comparison with the millisecond-level accuracy in NTP. Since advanced wireless research platforms enable experimentation in next-generation communication systems requiring fine-grained synchronization, PTP is considered as the basis for system-wide synchronization.  

In this paper, we present the design and implementation of PTP across the ARA wireless living lab~\cite{islam2024design,zhang2022ara}, i.e., \textit{AraSync}. We discuss in detail the ARA network topology from the synchronization perspective, PTP hardware and software architecture, and the configuration considerations for the devices involved in synchronization. To illustrate the characteristics and research capabilities enabled by \textit{AraSync}, we examine the behavior of PTP in ARA through extensive experimentation, measurement, and analysis. In our study, we include essential PTP metrics, such as offset from the master clock and Mean Path Delay~(MPD), at various levels of PTP  over both wired and wireless media and present a comparative analysis. Since long-range wireless links form the most feasible solution to ensure affordable broadband in rural areas, where fiber connectivity is scarce or absent, we measure the effectiveness of PTP over the 10.15\,km long \textit{AraHaul millimeter-wave} and \emph{microwave} links, especially the impact of weather conditions on PTP in synchronizing devices over wireless x-haul. It is observed that the accuracy of the synchronization varies more during high rain rates compared to moderate or light rain conditions. In short, PTP over long-range \textit{AraHaul}~\cite{10520543} wireless links achieves sub-microsecond level accuracy that is similar to its wired counterpart, however incurring more fluctuations.

The rest of the paper is organized as follows. Section~\ref{sec:PTP} provides an overview of the PTP operations and mechanisms. In Section~\ref{sec:arasync_framework}, we discuss the PTP network topology for \textit{AraSync}, including the capabilities and configurations of PTP-supported devices. Section~\ref{sec:performance_analysis} discusses our experiments, analysis, and observations. Finally, we conclude the paper in Section~\ref{sec:conclusion} with a discussion on the potential utility of \textit{AraSync} in future research work.

\section{Related Work}
\label{sec:related_work}

Precision Time Protocol~(PTP) has been extensively used for synchronizing nodes in time-sensitive systems.  In fact, the use of PTP is inevitable in experimental platforms such as ARA \cite{zhang2022ara}, enabling high-precision measurement studies in time-sensitive communication frameworks using 5G/6G platforms such as OpenAirInterface and Open RAN. 

In~\cite{7449285}, authors explored the fundamental principles of the PTP and shared their experience of its implementation in power system networks to realize sub-microsecond-level accuracy. However, the experiments were confined to a small laboratory setup rather than a real-world environment. Moreover, instead of an external grand-master clock, the authors used the Oven Controlled Crystal Oscillator~(OCXO) of the master clock as the time source.  As far as the wireless networks are concerned, authors of~\cite{8171749} proposed the Wireless Precision Time Protocol~(WPTP), an enhanced time synchronization protocol designed for over-the-air use. WPTP minimizes the number of PTP packet transmissions by consolidating timestamp information from \textit{Sync} and \textit{Delay\_Request} messages into \textit{DRPLY} message, thereby reducing the combined quantity of event and general PTP messages. Simulations demonstrated the effectiveness of WPTP in reducing network convergence time in a wireless multi-hop network. 

From the perspective of real-world implementation of PTP, authors of~\cite{10294864} detailed the step-by-step installation of \textit{LinuxPTP} on a testbed built on a 64-bit Raspbian operating system. From the measurements, it was observed that the TimeTransmitter or Master’s offset had a median value of 0\,ns with a standard deviation of 55\,ns. In~\cite{10296989}, authors proposed Simple PTP~(SPTP), a streamlined implementation of the existing PTP using the unicast UDP profile. SPTP simplifies the two-step PTP clock mode by eliminating the handshake mechanism and transferring the packet exchange control from the master to the client/slave. Subsequently, SPTP operates in a single-step mode without slave-master negotiation and the message exchanges are driven by the slave, thereby resulting in reduced network overhead.

It is important to note that fine-grained synchronization is a critical component of any 5G/6G 
communication systems. For instance, the crucial role of PTP-based synchronization in the emerging Open RAN systems is highlighted in~\cite{10559932}. A dedicated synchronization plane (S-plane) in Open RAN is responsible for synchronization among different components to ensure timely scheduling information to ensure proper Carrier Aggregation~(CA), avoid interference, and prevent call dropouts. Open fronthaul requires precise timing accuracy with a maximum allowable time error of 3\,$\mu$s. Besides synchronization, the authors discussed possible security threats in the S-plane and proposed effective countermeasures to address these challenges. In this paper, we discuss the design and implementation 
of PTP-based synchronization in the ARA wireless living lab involving long-range, high-capacity wireless x-hauls 
for enabling advanced wireless research.

\section{Precision Time Protocol}
\label{sec:PTP}
In this section, we present a brief overview of PTP, including the types of clocks and the delay mechanism. 

\subsection{PTP Clock Types and States}
\label{subsec:ptp_clock_types}

PTP uses four different types of clocks~\cite {PTPTrans76online,Configur56online} as well as different kind of messages between them for synchronizing the nodes. Different types of clocks include:
\begin{enumerate}
    \item \textbf{Grandmaster Clock~(GMC)}: The clock has a high-precision oscillator with capability to synchronize it from a GNSS receiver. In PTP, the GMC always acts as the \textit{master} clock.

    \item \textbf{Boundary Clock~(BC):} Acts as the second level of clock and, serves as the \textit{ordinary} clock for the GMC and \textit{master} clock for the third level \textit{ordinary} clocks. BCs have multiple PTP ports, each capable of establishing distinct paths to maintain the PTP clock hierarchy. 

    \item \textbf{Slave or Ordinary Clock~(OC):} Always the last level nodes (i.e., the devices that require synchronization) in the network hierarchy. OC has a single port which receives time from the \textit{master}. 
    
    \item \textbf{Transparent Clock~(TC): } The function of a TC device is to forward messages while keeping track of the residence time  of each message across the device (i.e., time taken for the message to traverse through the device from an ingress port to an egress port), allowing for the total time to be corrected at the destination. A TC has two modes: end-to-end mode and peer-to-peer mode. The end-to-end TC uses the end-to-end delay measurement mechanism between slave clocks and the master clock. The peer-to-peer TC involves link delay correction using peer-to-peer delay measurement mechanism, in addition to the residence time correction.
\end{enumerate}

On setting up the above-mentioned clocks at different levels, PTP achieves synchronization through exchanging messages of the following categories~\cite{9456762,Arnold2023May}:
\begin{enumerate}
    \item \textbf{Event messages: } These messages are tagged with timestamps when data packets arrive or leave a port and are used to calculate the link-delay based on these timestamps. They are time-critical since the accuracy in the transmit and receive timestamping directly affects the clock distribution accuracy. Event messages include \emph{Sync}, \emph{Delay\_Request}, \emph{Pdelay\_Request}, and \emph{Pdelay\_Response} messages.

    \item \textbf{General messages:} These messages are more conventional protocol data units, and the data in these messages is of importance to PTP, but their transmit and receive timestamps are not as important. They are used for purposes such as establishing Master-Slave hierarchies, and they include \emph{Announce}, \emph{Follow\_Up}, \emph{Delay\_Response}, \emph{Pdelay\_Response\_Follow\_Up}, \emph{Management}, and \emph{Signaling} messages. 
    
\end{enumerate}

The ports associated with each clock in PTP can be in any of the two states: (1)~\textit{master} or (2)~\textit{slave}. In fact, the port state primarily depends on the BC~\cite{9456762,10070440}. The \textit{master} port synchronizes the clocks connected downstream, while the \textit{slave} port synchronizes with the upstream \textit{master} clock. A unique state, called the \textit{passive} state, occurs when a BC receives PTP traffic from two or more different \textit{master} clocks, and the \textit{slave} clock selects only one \textit{slave} port for its synchronization using the Best Master Clock Algorithm~(BMCA) or Best TimeTransmitter Clock Algorithm~(BTCA)~\cite{10070440}. 

BMCA is used to decide port states, \textit{master} or \textit{slave}~\cite{9120376}, to form a Directed Acyclic Graph (DAG) for disseminating GMC time. When a clock is up, it enters the listening state to recieve the \emph{Sync} messages from a \textit{master} clock. Based on the received \emph{Sync} message and the clock properties (such as the administrator-assigned clock priority, clock class, and clock precision), the clock adopts its \textit{master} or \textit{slave} role.
When multiple BCs interconnect and form a cycle, certain ports are turned to \textit{passive} state to break the cycle, thereby maintaining the DAG structure~\cite{9456762}. TCs are not involved in the execution of BMCA, but they forward the PTP messages between the associated \emph{master} clocks and \emph{slave} clocks by adding the residence time. 

\subsection{Delay Measurement Mechanism}
\label{subsec:delay_mechanism}

\begin{figure}[!htbp]
        \centering
        \includegraphics[width=0.99\columnwidth]{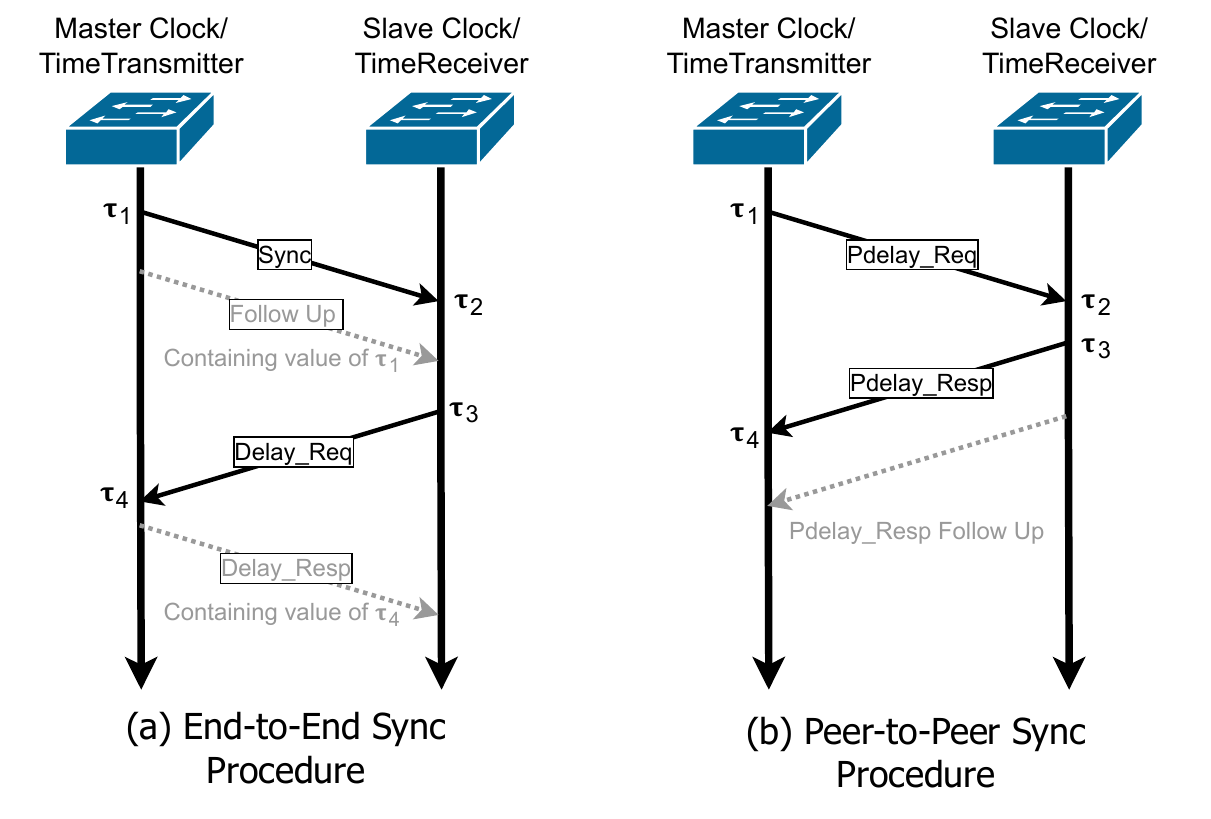}
        \caption{Delay measurement mechanisms.\vspace{-7mm}}
        \label{fig:Delay_mecha}
    \end{figure}

To synchronize between a pair of clocks, PTP adopts two mechanisms, i.e., end-to-end~\cite{9120376, 9121845} and point-to-point~\cite{9120376,Cisco_PTP_L2}, by measuring the delay between the clocks, as shown in Figure~\ref{fig:Delay_mecha}.  They are collectively referred to as Delay Measurement Mechanisms.

\paragraph{End-to-End Delay Measurement:}
\textit{Master} clocks (or TimeTransmitters) continuously send \emph{Sync} messages to all their \textit{slave} clocks (or TimeReceivers) at specified intervals. The \textit{master} clock records the sending time $\tau_1$ and \textit{slaves} record the reception time $\tau_2$ and, the time difference between them becomes $\tau_2 - \tau_1$. The \textit{master} communicates $\tau_1$ to the \textit{slave} using \textit{Sync} message itself or using a separate \textit{Follow Up} message. Given that the time difference between \textit{master} and \textit{slave} is the sum of their clock offset~($\Delta$) and the delay~$(\delta$) in message transmission, i.e., $\Delta + \delta$, we have $\tau_2 - \tau_1 = \Delta + \delta$. 
    The \textit{slave} clock, unaware of the delay, periodically sends \emph{Delay Request} messages to the \textit{master} clock, noting the sending time~($\tau_3$). The \textit{master} receives this message, records the arrival time ($\tau_4$), and responds to \textit{slave} with a \emph{Delay Response} message containing $\tau_4$. The time difference between the \textit{slave} and \textit{master} is $\tau_4 - \tau_3$, 
    and for the \emph{Delay Request} message, it can be expressed as -$\Delta$ + $\delta$; that is, $\tau_4 - \tau_3 = -\Delta + \delta$.
Assuming the delay is the same in both directions, the offset and Mean Path Delay~(MPD) can be computed as $\Delta = \frac{(\tau_{2}-\tau_{1}) - (\tau_{4}-\tau_{3})}{2}$ and $\delta = \frac{(\tau_{2}-\tau_{1}) + (\tau_{4}-\tau_{3})}{2}$, respectively~\cite{9121845}. 

\begin{figure*}[t!]
  \centering
  \includegraphics[width=\textwidth]{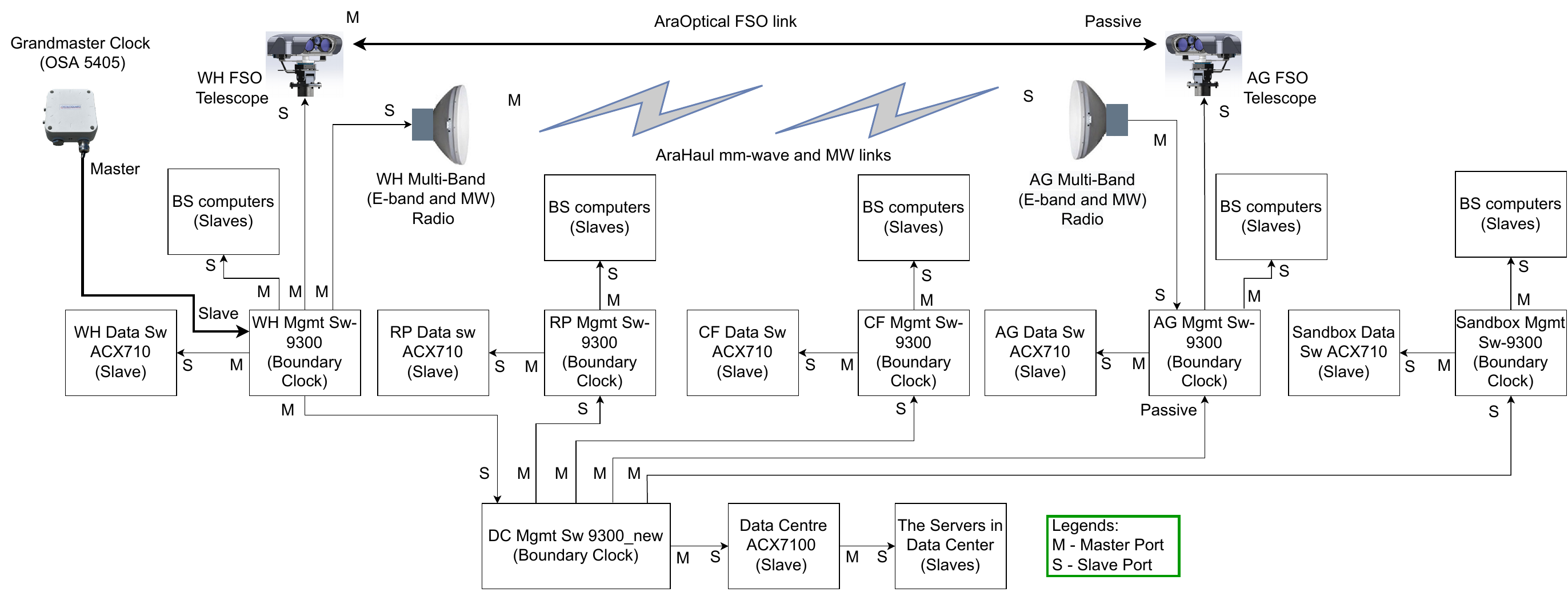}
  \caption{Network topology for \textit{AraSync}.\vspace{-5mm}}
  \label{fig:ARA time sync}
\end{figure*}

\paragraph{Peer-to-Peer Delay Measurement:} In Ethernet networks, the assumption that round-trip time~(RTT) being $\frac{\delta}{2}$ may not always be valid. To this end, peer-to-peer TCs measure the link peer delay to improve PTP accuracy.  
    Unlike the end-to-end approach, peer-to-peer synchronization allows PTP devices to exchange peer-delay measurement messages with their direct neighbors. Periodically, each device initiates the peer-delay exchange and adjusts it by updating the correction field in either the \emph{Sync} or \emph{Follow Up} message as it enters the device. Peer link delay can be expressed as $\delta_{link} = \frac{(\tau_{2}-\tau_{1}) + (\tau_{4}-\tau_{3})}{2}$, 
    where $\tau_{1}$ and $\tau_{2}$ are the timestamps at the \textit{master} and \textit{slave} for \emph{Pdelay Request} message, respectively, and $\tau_{3}$ and $\tau_{4}$ are the departure time at the \textit{slave} and arrival time at the \textit{master}, respectively.

\section{AraSync Framework}
\label{sec:arasync_framework}
In ARA~\cite{islam2024design}, we implemented PTP across the entire network using a tree/hierarchical topology~\cite{9456762} where the boundary clocks form the branch points in the tree depicted in Figure~\ref{fig:ARA time sync}. A single outdoor grandmaster clock~(GMC) is used to synchronize the network. The GMC is a smart Global Navigation Satellite System~(GNSS) receiver leveraging signals from available satellites for synchronization. The GMC connects to a PTP-enabled switch, serving as the first-level BC in the hierarchical topology and acting as the \textit{master} for all the downstream devices. The first-level BC forwards the PTP traffic through wireless x-haul and fiber links to the second-level BCs via Layer 2 Multicast. The second-level BCs then act as \textit{masters} for their respective downstream devices forwarding PTP messages via Layer 2 multicast. Further, the second-level BCs act as the synchronization sources to the third-level BCs and the devices connected to them. In short, for \textit{AraSync}, we use a 3-level hierarchical topology for synchronization. 

Since the BC is a switch with a large number of ports, the tree topology allows it to serve as the \textit{master} to a large number of OCs connected to it. The first-level BC becomes a \textit{slave} only to the GMC through the port where GMC is connected, whereas the remaining ports act as the \textit{master} for the devices connected to them. The second-level BCs serve as \textit{slave} to the first-level BC and act as \textit{master} for any devices connected to them. The switches connected directly with the second-level BC act as the last-level BCs and provide time synchronization to the end devices, i.e., the OCs. In case a BC receives PTP \emph{Event} and \emph{General messages} from two higher-level BCs, the BC with fewer number of hops toward GMC is selected as the \textit{master} in accordance with the Best Master Clock Algorithm~(BMCA). Consequently, the port of the downstream device facing the BC with a lower hop-count will be configured as the \textit{slave}, while the port connected to the BC with a higher hop-count is set to the \textit{passive} state. For instance, one of the base station switches, i.e., Agronomy Farm Cisco Catalyst C9300, has wireless x-haul connectivity with the first-level BC (Wilson Hall Cisco Catalyst C9300) and fiber connectivity with the second-level BC (Data Center Cisco Catalyst C9300), illustrated in Figure~\ref{fig:ARA time sync}. The wireless x-haul link spans approximately 6.31\,miles between the two base stations, whereas the optical fiber link extends nearly 7.5\,miles. The BMCA sets the port connecting first level BC (i.e., the one via wireless x-haul) as the \textit{slave} due to the lower hop-count from GMC. On the other hand, the port connected to the fiber is set to \textit{passive} state due to higher hop-count and pruned to avoid the possibility of cycle in the topology. In short, the BMCA determines the states of the PTP ports and constructs the synchronization tree.

\subsection{PTP Devices in ARA}
\label{subsec:ptp_devices_ara}

\begin{figure}[t!]
          \centering
          \includegraphics[width=0.65\columnwidth]{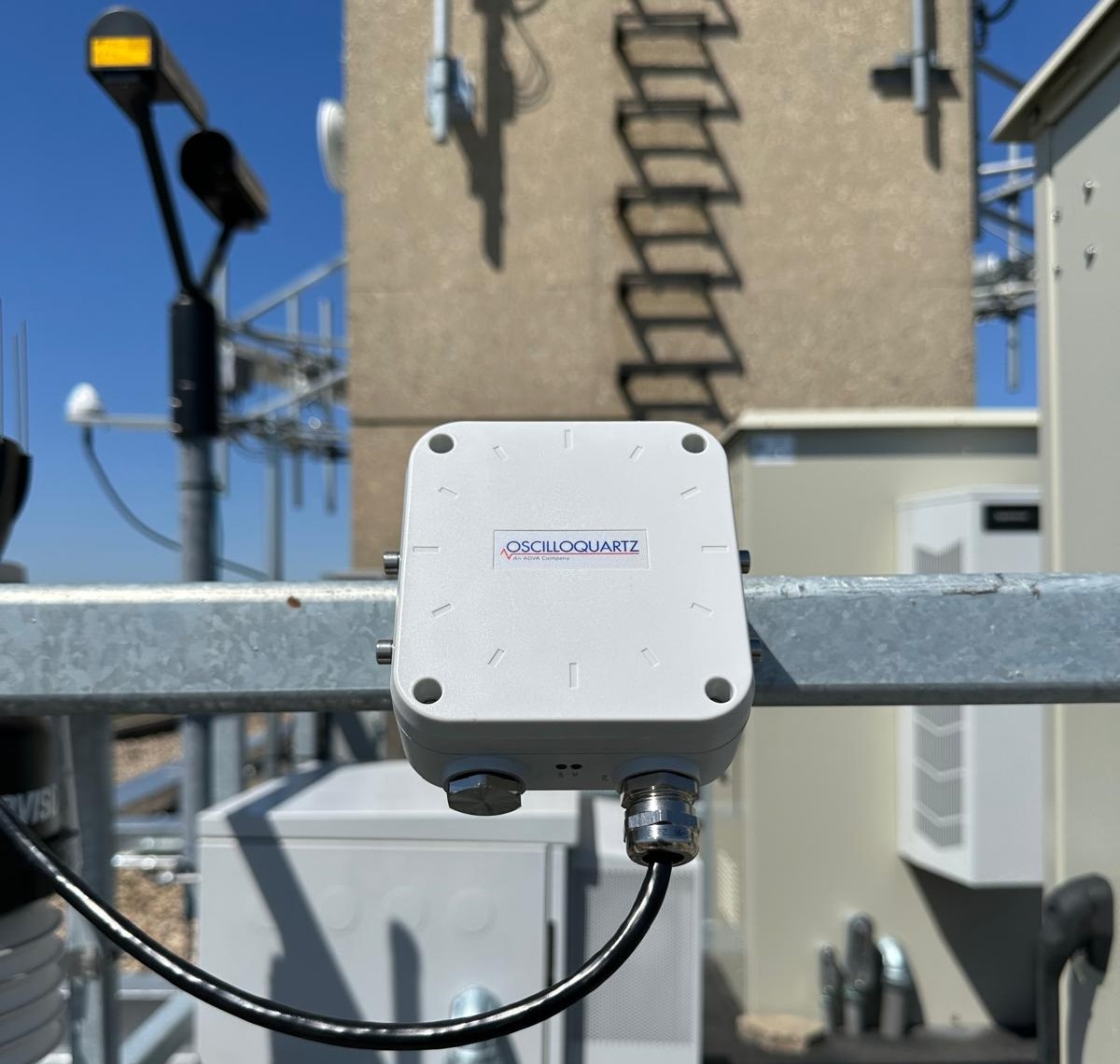}
          \caption{\emph{AraSync} GMC at the Wilson Hall rooftop.}
          \label{fig:ARA_GMC}
        \end{figure}

The Oscilloquartz OSA 5405 Outdoor SyncReach device, a GNSS antenna integrated with GNSS receivers and PTP stacks, is stationed at the Wilson Hall base station site (see Figure~\ref{fig:ARA_GMC}) as the GMC to serve as a precise time source~\cite{OSA5405S41:online}. The OSA 5405 GMC supports IEEE 1588 2008 L3/L2 and ITU-T 8265.1/8275.1/8275.2/power/broadcast/enterprise standards. It facilitates PTP over both IPv4 and IPv6 simultaneously and includes fallback options for PTP and Synchronous Ethernet (SyncE) inputs. In addition, the GMC supports both multicast and unicast modes. 
        
The Cisco Catalyst C9300 switches function as BC at various levels of \emph{AraSync} serving the downstream devices. These switches support both Layer-2 and Layer-3 multicast PTP synchronization protocols, specifically conforming to the IEEE 1588/8275.1 profile, and enable precise clock synchronization with sub-microsecond accuracy~\cite{CiscoCat48:online}. The Juniper ACX7100 Cloud Metro Router~\cite{acx7100c90:online} and ACX710 Universal Metro Router~\cite{acx710un41:online}, which serve as the data traffic carriers, function as end-devices in \textit{AraSync}. These routers support Layer-3 unicast PTP synchronization mode, exclusively adhering to the g.8275.2 profile, as well as SyncE. The integration of PTP and SyncE synchronization is referred to as Hybrid Mode.

The compute nodes at the base station, data center, and sandbox in ARA are Dell R750 PowerEdge servers with Intel X710 SFP+ Network Interface Cards~(NICs) and Broadcom 5720 1\,Gb Ethernet NIC, which support hardware and software timestamping required for PTP. The Ethernet port is used for the management network, including PTP sync, while the SFP+ ports are for the data network. The compute servers run Ubuntu 20.04~LTS with \textit{LinuxPTP}~\cite{10294864} installed and configured on the servers for the PTP time Sync. \textit{LinuxPTP} provides \emph{ptp4l}, the daemon that synchronizes PTP Hardware Clock~(PHC) from the NIC, and \emph{phc2sys},  which synchronizes PHC and the system clock. \textit{LinuxPTP} synchronize the servers to the respective switches that they are connected to. The compute nodes at the User Equipment~(UE) have a high-precision GPS sensor, and the GPS data received by the sensor is fed as the source for NTP synchronization. We use the \textit{gpsd} daemon to periodically poll the GPS signals.

\subsection{PTP Configuration in ARA}

The GMC is configured as Layer-2 PTP multicast to forward the PTP traffic to the downstream PTP instances. The PTP profile is configured as g8275.1, with the domain number being 24, the lowest domain number in the domain range of 24--43 for this PTP profile. The \textit{Announce, Sync,} and \textit{Delay Response} rates are set at 8~pps,16~pps, and 16~pps, respectively, to send out the PTP messages in a very frequent fashion, and these rates are the maximum as per the standard. The \textit{Sync} messages are sent in one-step mode, i.e., the \textit{Sync} messages contain the initial timestamp to the TimeReceiver instances. The Global Navigation Satellite System~(GNSS) is the primary way to synchronize the OSA-5405 GMC in GPS mode and tracking mode is enabled with the GNSS-tracking system. The GMC also provides SNR, elevation values, latitude, longitude, altitude, and the available satellites, from where the GMC receives the GPS signals for synchronization. Similar to the GMC, the Cisco Catalyst C9300 switches are configured to Layer-2 PTP multicast. As IEEE1588/8275.1 is the only profile available in Cisco C9300, we set the domain number to~24. The \textit{Announce} message interval is 3~(i.e., 8 pps) while the \textit{Sync} interval is -4~(i.e., 1/16 pps). We configure the end-to-end delay mechanism to compute the offset and mean path delay, and the mode of the switch is configured as a boundary clock delay request.

\begin{figure}[b!]
    \centering
    \includegraphics[width=1\columnwidth]{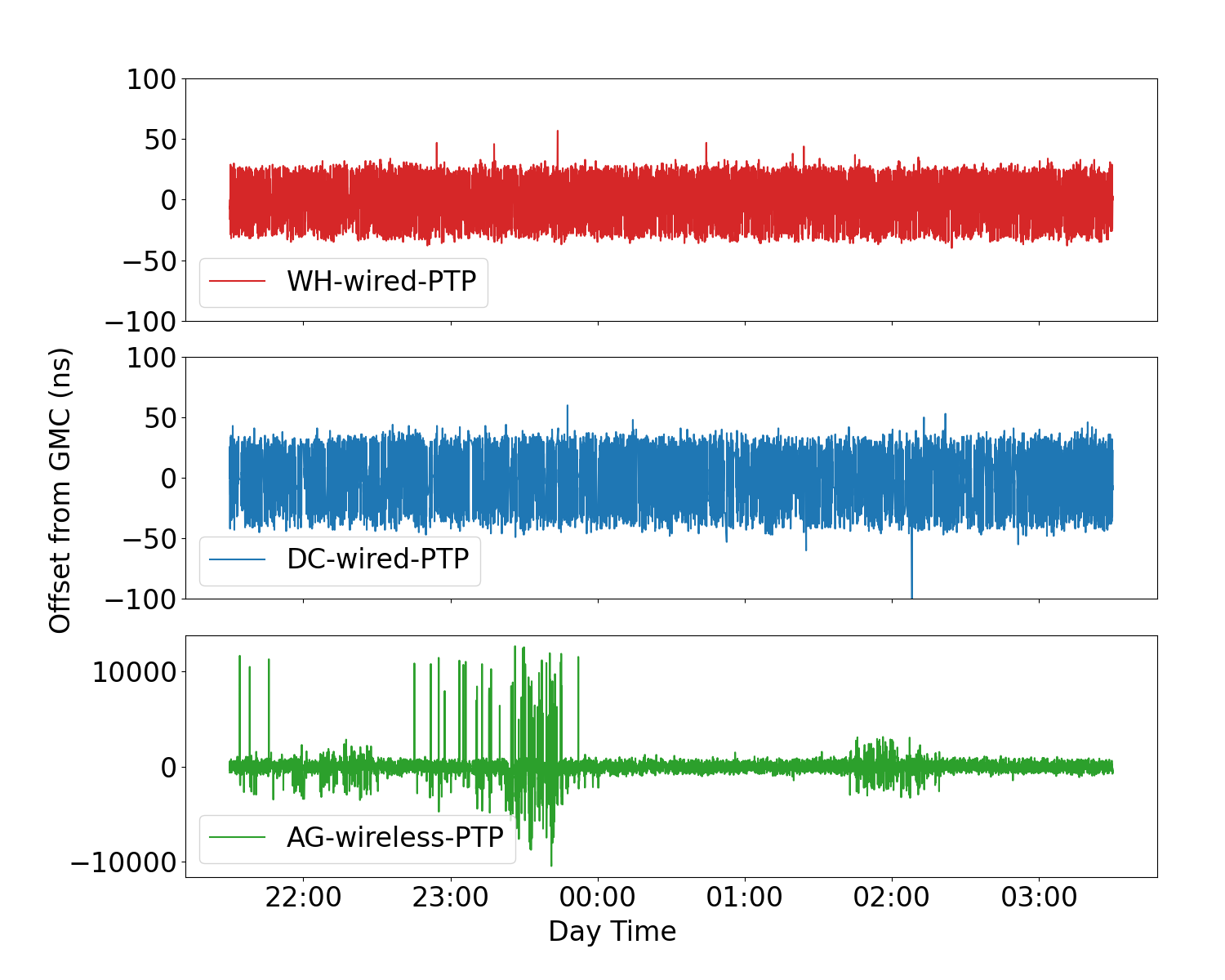}
    \caption{\textit{Wireless}- and \textit{wired}-PTP offset from GMC at different levels of BCs.}
    \label{fig:PTP_WH_DC_AG_aviat_072724_2_Offset_Stacked}
\end{figure}

\section{Performance Analysis}
\label{sec:performance_analysis}

For analyzing the performance of \textit{AraSync}, we first consider the offset of BCs at different levels with the GMC. For our analysis, we consider Wilson Hall Cisco C9300 as the first-level BC~(connected directly to GMC), Data Center Cisco C9300 as the second-level \textit{wired} BC~(connected to Wilson Hall C9300 via Fiber cable), and Agronomy Farm Cisco C9300 as the second-level \textit{wireless} BC~(connected to Wilson Hall C9300 via the AraHaul link). The behavior of offset observed is shown in Figure~\ref{fig:PTP_WH_DC_AG_aviat_072724_2_Offset_Stacked}. The range of offset from GMC to first-level BC (Wilson Hall Cisco C9300 switch) varies from -25\,ns to 25\,ns. The second-level BC, which is connected to the first-level BC via fiber, provides an offset from -45\,ns to 45\,ns, i.e., 20\,ns higher than first-level BC. However, the offset measured at the Agronomy Farm BC fluctuates between -10,000\,ns to 10,000\,ns. This second-level BC relies on a first-level BC, which transmits PTP messages over \textit{AraHaul}, and fluctuations in offset can be attributed to the wireless link's channel health and varying weather conditions.

\begin{figure}[t!]
    \centering
        \includegraphics[width=.95\columnwidth]{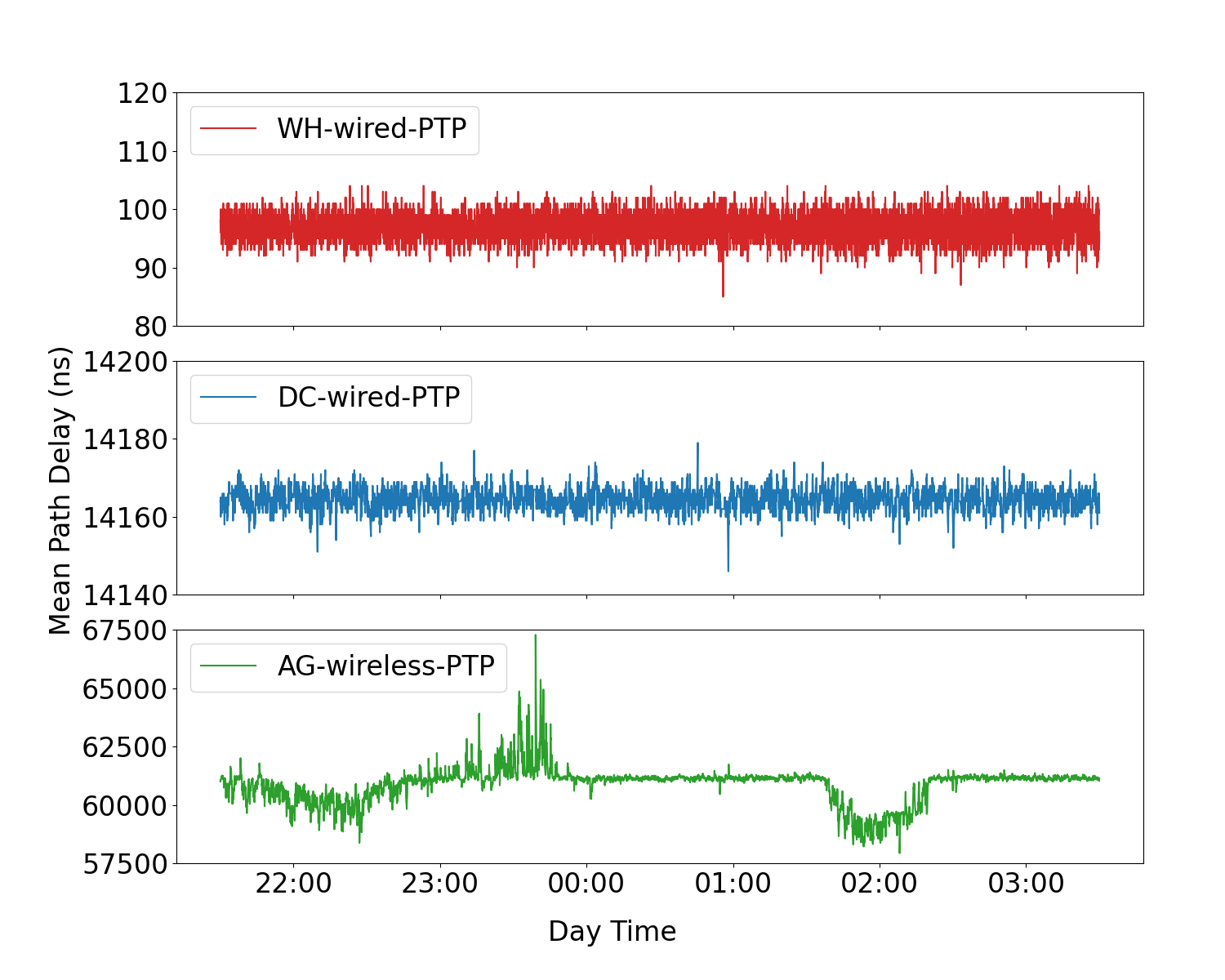}
        \caption{Mean Path Delays~(MPDs) of \textit{wireless}- and \textit{wired}-PTP at different levels of BCs.}
        \label{fig:PTP_WH_DC_AG_aviat_072724_2_MPD_Stacked}
\end{figure}

In Figure~\ref{fig:PTP_WH_DC_AG_aviat_072724_2_MPD_Stacked}, the Mean Path Delay~(MPD) between the first-level BC, i.e., Wilson Hall, and the second-level \textit{wired}-PTP BC (i.e., Data Center) is approximately 14,000~ns, which reflects the delay introduced during PTP message transmission via optical fiber. Conversely, the second-level \textit{wireless}-PTP BC at Agronomy Farm shows a delay of nearly 60,000 ns, adding an extra 46,000 ns compared to the fiber medium.

\begin{figure}[t!]
    \centering
        \includegraphics[width=.91\columnwidth]{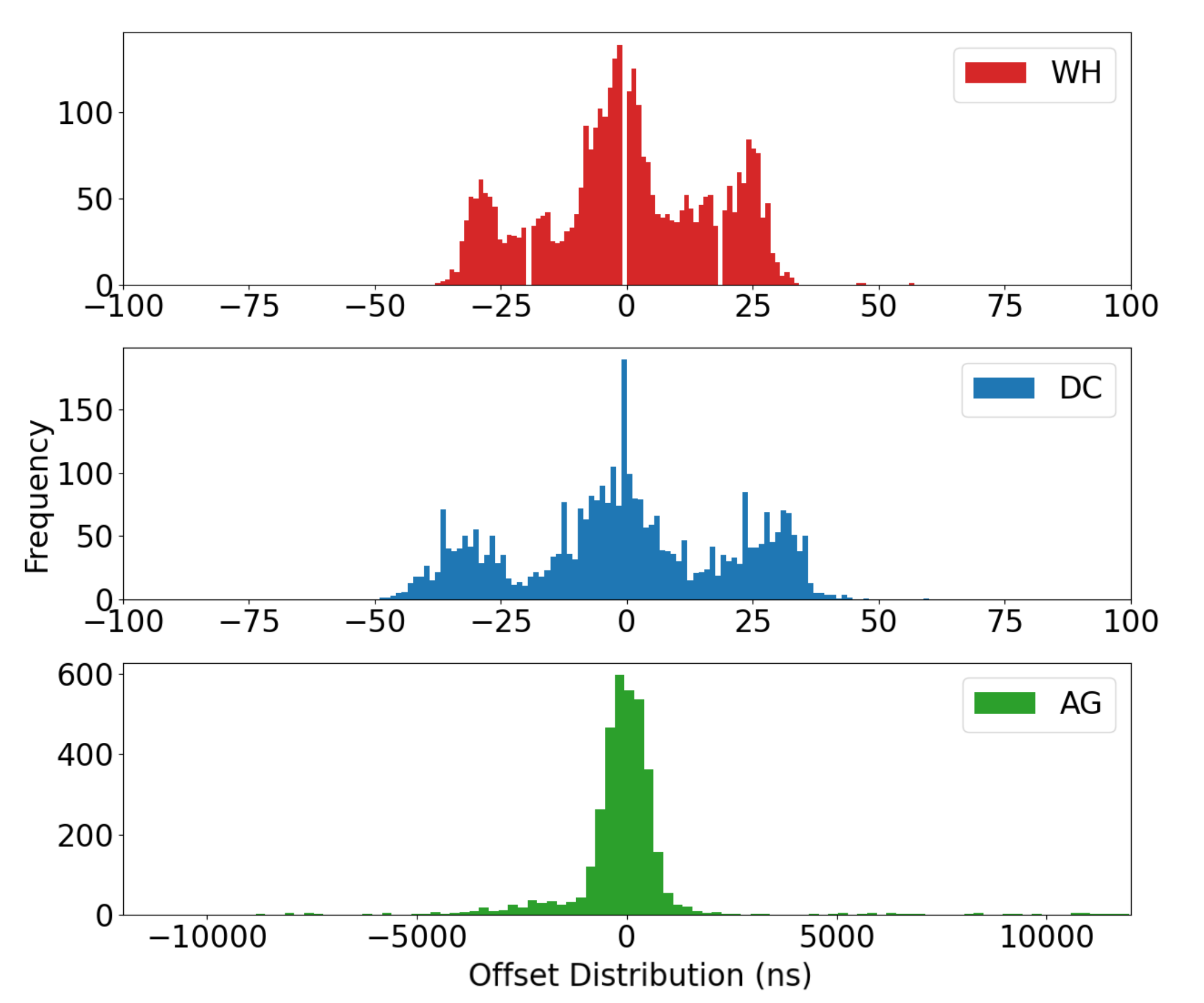}
        \caption{Offset distribution between \textit{wireless}-PTP and \textit{wired}-PTP at different levels of BCs.}
        \label{fig:PTP_WH_DC_AG_aviat_072724_2_Histograms}
\end{figure}

The distribution of offset of \textit{wireless}-PTP and \textit{wired}-PTP at different levels of BC is illustrated in Figure~\ref{fig:PTP_WH_DC_AG_aviat_072724_2_Histograms}. 
At the first-level BC (Wilson Hall), the offset for \textit{wired}-PTP is tightly clustered around 0\,ns, ranging from -35\,ns to 35\,ns.  
At the second-level BC (Data Center), the \textit{wired}-PTP distribution remains stable, ranging from -50 to 50 ns, showing a minimal increase in time error. Conversely, the second-level \textit{wireless}-PTP BC at Agronomy Farm shows a much broader distribution of time error, ranging from -6,000\,ns to 6,000\,ns. This significant increase in offset can be attributed to the inherent variability and less predictable nature of wireless channels as well as the impact of environmental factors on channel conditions. Figure~\ref{fig:PTP_WH_DC_AG_aviat_072724_2_Histograms}  highlights the challenges associated with wireless-PTP propagation, where environmental factors and the less predictable nature of wireless channels lead to synchronization discrepancies. On the contrary, wired-PTP shows greater stability and lower time error associated with propagation across different levels of BCs.

\begin{figure}[htbp!]
    \centering
        \includegraphics[width=1\columnwidth]{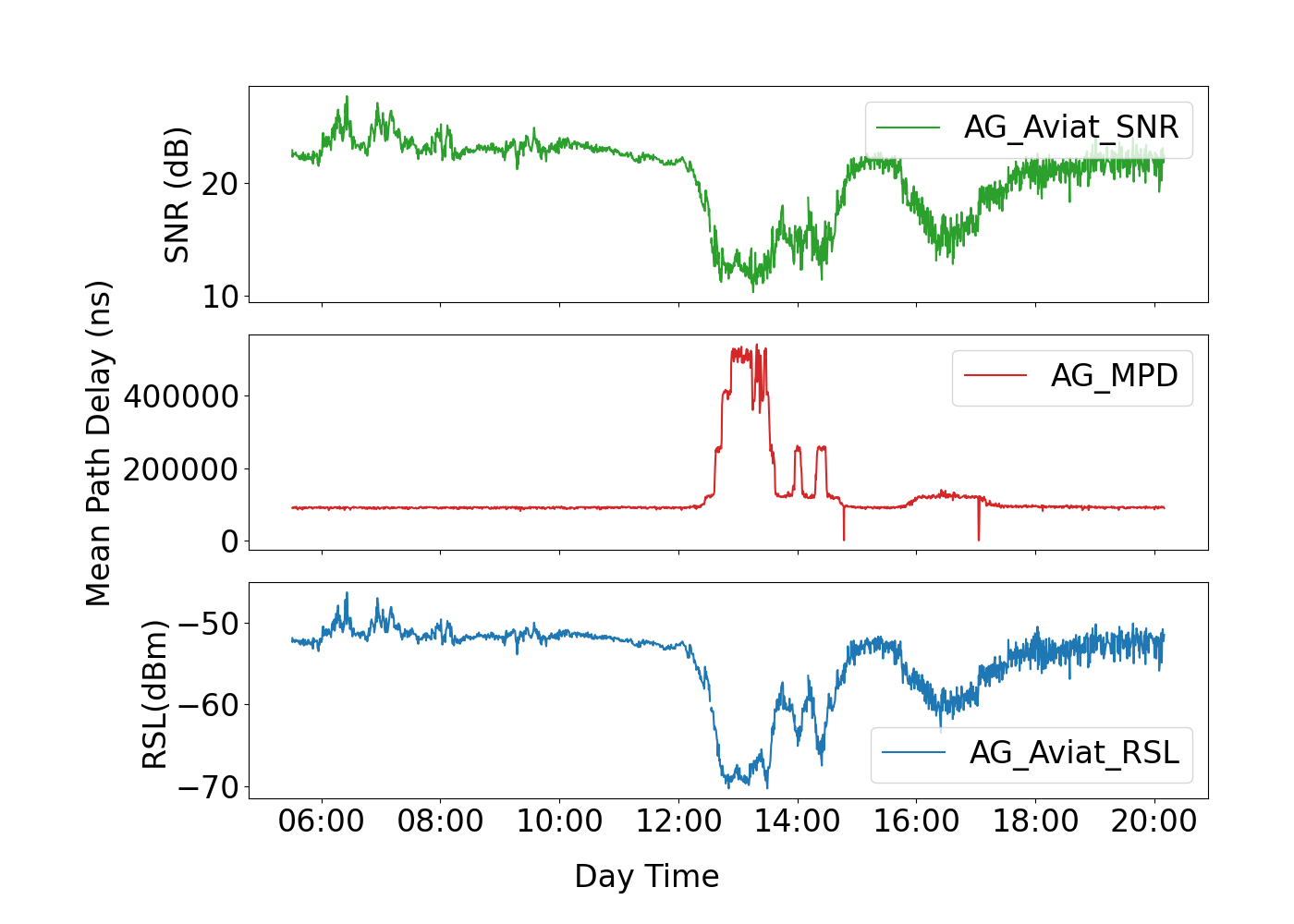}
        \caption{\textit{Wireless}-PTP Mean Path Delay dependencies on wireless channel health.
        \vspace{-6mm}
        }
        \label{fig:PTP_AG_aviat_snr_MPD_rsl_072924}
\end{figure}

The dependencies of \textit{wireless}-PTP Mean Path Delay~(MPD) on the health parameters of wireless link, such as Signal-to-Noise Ratio (SNR) and Received Signal Level~(RSL), are illustrated in Figure~\ref{fig:PTP_AG_aviat_snr_MPD_rsl_072924}. 
The Spearman correlation coefficient of MPD with RSL is -0.738, while with SNR, the coefficient is -0.754, indicating strong negative monotonic dependencies. The negative correlation coefficients underscore the sensitivity of PTP performance to wireless channel conditions. Figure~\ref{fig:PTP_AG_aviat_snr_MPD_rsl_072924} shows the characteristics of wireless channels for the duration of 14~hours. 
It is observed that higher SNR values, indicating better signal quality, correspond to lower MPD values. 
The RSL over the same period indicates a strong negative correlation with MPD. 

\begin{figure}[t!]
    \centering
        \includegraphics[width=.96\columnwidth]{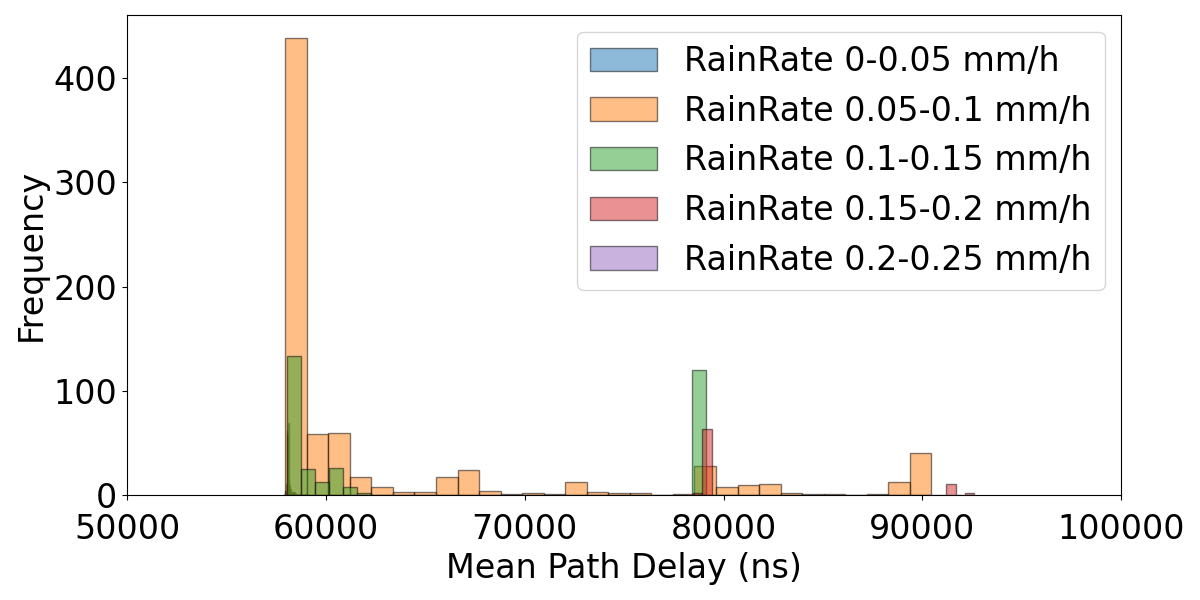}
        \caption{\textit{Wireless}-PTP: MPD dependencies on rain-rate.}
        \label{fig:PTP_AG_MPD_rainrate_073024}
\end{figure}

Figure~\ref{fig:PTP_AG_MPD_rainrate_073024} indicates the distribution of MPD under varying rain conditions.
The MPDs are predominantly clustered around 60,000\,ns for rain rates of 0--0.05\,mm/h and 0.05--0.1\,mm/h, with a particularly high frequency for the latter. Another notable cluster is observed around 80,000\,ns for the rain rates of 0.1--0.15\,mm/h, 0.15--0.2\,mm/h, and 0.2--0.25\,mm/h. It is interesting to note that higher rain rates result in distributed and less frequent MPD values indicating more variability and less predictable PTP transmission, potentially leading to higher jitters and packet delays. In contrast, lower rain rates are associated with more frequent and stable MPD values, suggesting more consistent path delays.

\begin{figure}[t!]
    \centering
        \includegraphics[width=.95\columnwidth]{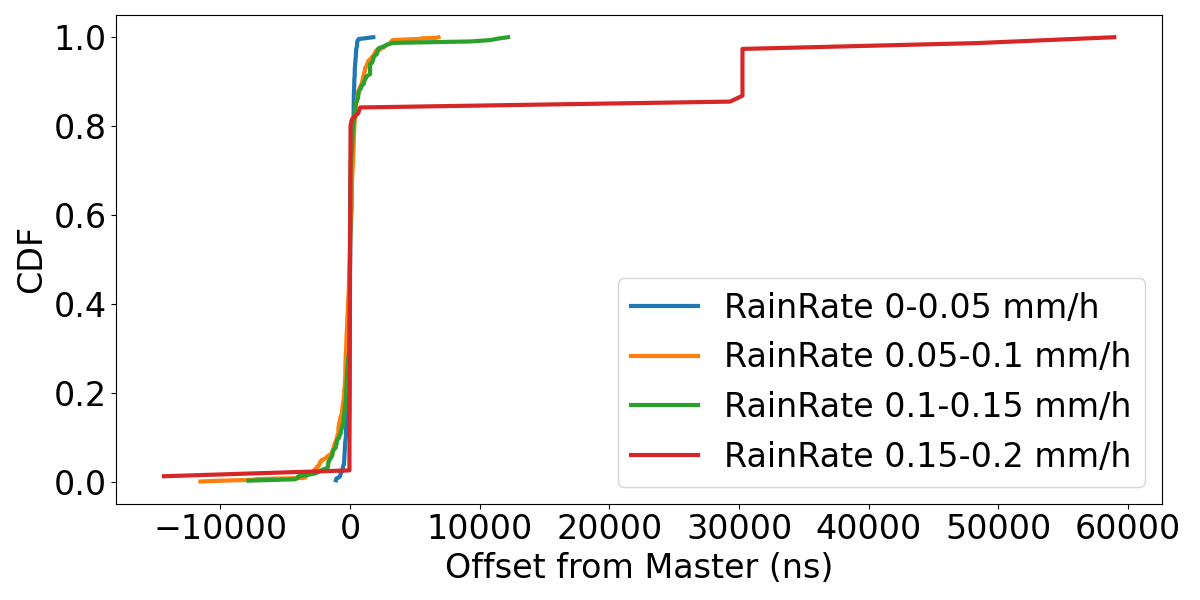}
        \caption{\textit{Wireless}-PTP offset from GMC vs. rain-rate.}
        \label{fig:PTP_AG_offset_rainrate_CDF_073024}
\end{figure}

In Figure~\ref{fig:PTP_AG_offset_rainrate_CDF_073024}, we show the Cumulative Distribution Function~(CDF) illustrating the impact of rain on the second-level BC offset, i.e., the Agronomy Farm clock connected over AraHaul mmWave wireless channel operates at 80\,GHz, from the GMC.
Low rain-rates (0--0.05\,mm/h, 0.05--0.1\,mm/h) result in most offsets close to zero, indicating minimal offset and high synchronization accuracy. For moderate rain (0.15--0.2\,mm/h) follows a similar trend as that of low rain-rate, however, with slightly more offset variability.  Higher rain-rates (0.15--0.2\,mm/h) show a gradual CDF increase, indicating more offset spread (up to 60,000\,ns) and higher variability, i.e., heavy rain negatively impacts the precision of PTP-based synchronization.

\begin{figure}[tb!]
    \centering
        \includegraphics[width=1\columnwidth]{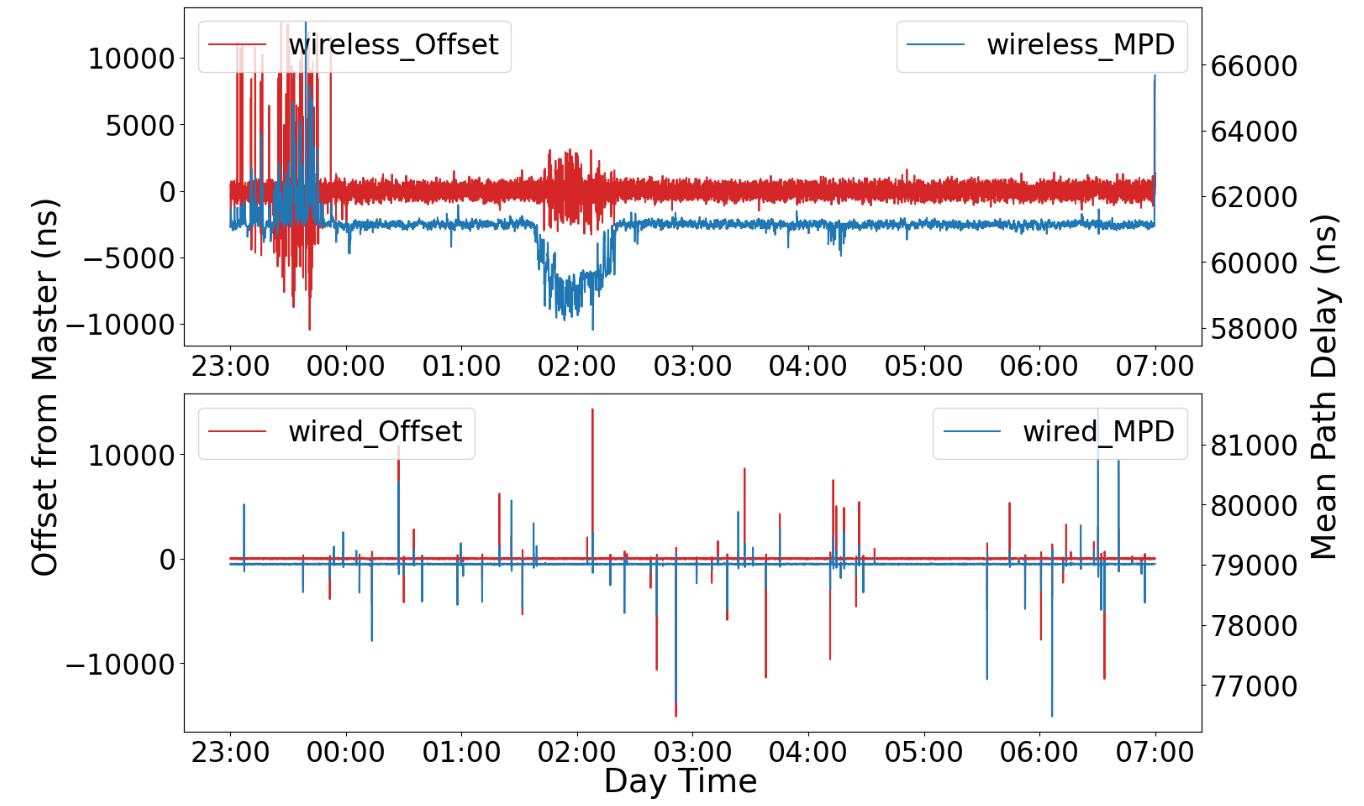}
        \vspace{.2mm}
        \caption{\textit{Wired}- and \textit{Wireless}-PTP offset and MPD from GMC at Agronomy Farm.}
        \label{fig:PTP_AG_aviat_072724_2_073124_MPD_Offset_Stacked2}
\end{figure}

The performance of PTP over wireless and wired connections at the Agronomy Farm BC is illustrated in Figure~\ref{fig:PTP_AG_aviat_072724_2_073124_MPD_Offset_Stacked2}.
\textit{Wired}-PTP shows significantly more stability in offset and MPD compared to \textit{wireless}-PTP, which experiences more fluctuations due to susceptibility to external factors, such as signal attenuation, interference, and environmental conditions, affecting synchronization precision.
Additionally, the \textit{wired}-PTP shows more consistent offset and MPD with fewer deviations, achieving better synchronization accuracy around 0\,ns offset. However, the MPD is higher due to higher hop counts. In contrast, \textit{wireless}-PTP has larger deviations, especially around 23:00 and 02:00, indicating less precise synchronization. These comparisons highlight the superior stability and consistency of \textit{wired}-PTP over \textit{wireless}-PTP. 
We could achieve sub-microsecond level PTP performance, nearly as good as \textit{wired}-PTP through the wireless link in this case; however, the \textit{wireless}-PTP fluctuation is higher due to external factors.

\vspace{.5mm}

\vspace{.1in}
\noindent \textbf{\textit{Lessons Learned: }}
The primary requirement to enable PTP synchronization is to select an optimal location for the GMC with sufficient satellite signal coverage. Since the data center cannot be considered the central point for \textit{AraSync} to place the GMC due to its enclosed nature and surrounding buildings, we chose the Wilson Hall rooftop for the GMC, ensuring a view of the unobstructed sky. Another major challenge includes the identification and resolution of the mismatch between PTP-profiles and the layer (L2 or L3) on which PTP operates in switches. For example, the GMC supports g8275.1 for L2-PTP and g8275.2 for L3-PTP, while the Cisco switch BC supports only g8275.1-L2/L3 PTP. In fact, on a PTP configuration change, mid-level BCs enter into an \textit{uncalibrated} state with their master, which was resolved by PTP rebooting multiple times on the first-level BC.

\vspace{.1in}
\noindent\textbf{\textit{AraSync Use Cases: }}
As an advanced research platform, PTP-based synchronization enables advanced wireless experiments in ARA involving time-critical network functions. For example, synchronization helps to precisely align data frames and time slots in 5G and beyond systems~\cite{pktr}
In fact, PTP is a critical requirement in Open~RAN to ensure the precise timing of the Enhanced Common Public Radio Interface~(eCPRI) interface between Distributed Unit~(DU) and Radio Unit~(RU). Further, \textit{AraSync} allows us to measure finer-level latency at different layers at both BS and UE to understand the impact of latency in URLLC applications. Finally, PTP is an essential requirement for massive MIMO and beamforming, network slicing, resource scheduling, and interference mitigation in Next-G wireless systems.

\section{Conclusion}
\label{sec:conclusion}

In this paper, we discuss \textit{AraSync}, an infrastructure-level framework for high-precision time synchronization to enable time-sensitive wireless experimentation. \textit{AraSync} is primarily built on the top of PTP with a grand master clock and different levels of boundary clocks. Even though PTP is used over fiber to synchronize large-scale systems, \textit{AraSync} is unique in synchronizing devices using PTP over long-range high-capacity wireless backhaul (i.e., \textit{AraHaul}) involving mmWave and microwave links in addition to the optical fiber based wired links. Further, we studied the variations in offset and mean path delay of BCs at different levels toward the GMC. Our measurements shed light on the impact of dynamic wireless channel characteristics as well as weather conditions on the synchronization accuracy. Our framework and analysis can help in designing time-sensitive next generation wireless communication systems, such as Open RAN, involving wireless x-haul. 

\balance

\bibliographystyle{ACM-Reference-Format}
\bibliography{ref}

\end{document}